\documentclass[12pt]{iopart}

\usepackage{graphicx} 
\begin{document}

\title[Continuous atom laser with BEC]{Continuous atom laser with Bose-Einstein condensates involving three-body interactions}
\author{A. V. Carpentier$^1$, H. Michinel$^1$, D. N. Olivieri$^2$ and D. Novoa$^1$}
\address{
$^1$ \'Area de \'Optica, Facultade de Ciencias de Ourense,
Universidade de Vigo, As Lagoas s/n, Ourense, ES-32004 Spain.\\
$^2$ \'Area de Linguaxes e sistemas inform\'aticos, Escola Superior de Enxe\~ner\'{\i}a Inform\'atica,
Universidade de Vigo, As Lagoas s/n, Ourense, ES-32004 Spain.}
\ead{avcarpentier@uvigo.es}

\begin{abstract}
We demonstrate, through numerical simulations, the emission of a coherent 
continuous matter wave of constant amplitude from a Bose-Einstein 
Condensate in a shallow optical dipole trap. The process is achieved by spatial control 
of the variations of the scattering length along the trapping axis, including elastic 
three body interactions due to dipole interactions. In our approach, the outcoupling 
mechanism are atomic interactions and thus, the trap remains unaltered. We 
calculate analytically the parameters for the experimental implementation of 
this CW atom laser.
\end{abstract}

\vspace{2pc}

\maketitle
\section{Introduction}  The experimental realization of Bose-Einstein condensates (BEC) with 
alkali atoms \cite{Anderson95} has opened the door to the realization of atomic beam sources 
of high coherence and brightness, with strong analogy to optical laser beams. Such ``atom lasers'' 
promise unprecedented achievements in fields such as atomic interferometry or gravimetry and have 
been proposed in several configurations, predominantly pulsed \cite{Mewes97,Carr04,rodas-verde05} 
or semicontinuous \cite{Bloch,Hagley,Bloch2,Pepe,Ultimo}, and with different outcoupling mechanisms
\cite{Guerin,Couvert,Gattobigio,Delgado}.

For instance, the first atom laser used short radio-frequency pulses to outcouple 
atoms from the cavity, flipping the spins of some of the atoms and  releasing them 
from the trap with the aid of gravity \cite{Mewes97}. However, despite its success 
in the first experiments on atom laser operation, spin-flipping techniques have 
some limitations as a versatile outcoupling mechanism for coherent matter-wave 
sources. In first place, there are strong fluctuations in the output at high fluxes 
due to the fact that the spin-flipping mechanism populates all accessible Zeeman 
states \cite{close}. This constitutes a strong drawback for practical applications 
of atom lasers in high precision measurements such as matter wave gyroscopes \cite{gustavson97}.
Moreover, in the former atom lasers, the emission process is driven by gravity and thus the
resulting matter wave is always emitted in the vertical direction. 

Atomic soliton lasers overcome the previous limitations using the mechanism of modulational 
instability to trigger the outcoupling of atoms from the cloud \cite{Carr04, rodas-verde05}. 
Thus, in this case there is no need for gravity since the emission is obtained by an adequate combination of nonlinear 
effects in the atom cloud and the manipulation of the trap.  In particular, the necessary condition is 
that the total number of particles in the cloud exceeds a critical
threshold.  Therefore, 
atomic soliton lasers add arbitrary directionality to the emitted beam. 
However, currently proposed matter-wave sources based on nonlinear effects can only operate 
in pulsed regime, due to modulational instability of the beam: a nonlinear effect which yields 
a burst of atomic solitons \cite{Perezgarcia98, solitons1,solitons2}. These {\em localized} nonlinear 
waves arise as a result of the perfect balance between dispersive and nonlinear effects in one
dimensional systems. A set of robust pulses that propagate without shape distortion is therefore
produced after the emission, thereby precluding continuous operation of the matter wave laser.

The goal of this paper is to present a novel outcoupling mechanism for atomic beam sources
which overcomes all the previous problems and produces directional continuous wave (CW) emission of an atomic
beam with constant amplitude. The trick is to use a combination of a shallow dipolar potential
with attractive two-body and repulsive three-body interactions. Such
interatomic potentials can be obtained for strongly dipolar atomic
systems in the presence of adequately tuned magnetic fields. \cite{zoller} 
We will show that this technique allows both the ability to extract atoms from the trap \cite{rodas-verde05} 
and to overcome modulational instability of the outcoupled beam, thus providing fluent CW operation of the 
matter-wave laser. Moreover, we will show that because of the inclusion of the three-body elastic repulsion
the emitted wave displays constant amplitude and is very robust against perturbations, permitting the
achievement of atomic coherent sources with unprecedented brightness and stability. 


\section{System configuration and theoretical model} Let us assume that an elongated 
BEC is strongly trapped in the transverse directions ($x,y$) by a parabolic potential 
$V_\perp$ and weakly confined in the longitudinal axis $z$ by a shallow dipole trap $V_z$
of width $L$ and depth $V_0$. We also assume a step variation along $z$ of the 
atomic interactions. Thus, for $z<0$ (where the condensate is initially 
placed at $t=0$) the interatomic forces are set to a negligible value whereas in the 
$z\geq 0$ region, the value of the scattering length is such as to yield strong two-body 
and three-body interactions comparable to the trapping force along $z$. 

We will consider the case of both two-body attractive and three-body repulsive forces 
in the region $z\geq 0$. The techniques for controlling the scattering
length which produces two-body attractive interactions are well-known and may include
optical or magnetic fields \cite{solitons2,FB1,FB2}. It has also been proposed that {\em repulsive}
three-body forces can be activated with an optical lattice driven by microwave fields in ultracold gases of
highly polar molecules \cite{zoller}, such as RbCs \cite{sage} or atoms like Cr \cite{griesmaier}.
For this technique, Hubbard models 
predict strong nearest-neighbor three-body interactions, whereas the two-body terms can be 
tuned independently. Also, recent experiments with ultracold
Cs atoms \cite{efimov} have revealed the existence of the so-called Efimov states, which represent 
a paradigm for universal quantum states in the three-body sector. These remarkable results
have opened the possibility to novel BEC applications based on three-body controlled 
interactions, as presented in this paper.

We will consider here a system of $N$ weakly interacting dipole bosons of mass $m$, 
trapped in a potential $V(\vec{r})$, where the evolution in time $t$ of the condensate wavefunction $\Psi$ 
is correctly described according to experiments \cite{giovanazzi} by a mean field 
Gross-Pitaevskii equation of the form
\begin{equation*}
\label{GPE}
i \hbar \frac{\partial \Psi}{\partial t} =
- \frac{\hbar^2}{2 m} \nabla^{2}\Psi +
V(\vec{r})\Psi + \Delta(z,|\Psi|^2)\Psi ,
\end{equation*}

where $N = \int |\Psi|^2 \ d^3 \mathbf{r}$.  The functions $V$ and $\Delta$ describe respectively 
the external trap and the dipole-dipole induced potential, which can be activated in the region $z>0$. 
In this work we consider a cigar-shaped BEC which is tightly confined in ($x,y$) by means of a tubular harmonic 
trap. A {\em shallow} potential provided by an optical dipole trap
$V_z$ \cite{Stamper98,Martikainen99} 
keeps the cloud trapped along the $z$ direction. Thus, the explicit form of the three-dimensional trap $V$ is given by:

\begin{equation*}
V(\vec{r})=V_\perp+V_z=\frac{m\nu^2_\perp}{2} 
\left( x^2+y^2 \right)+V_0(z/L),
\end{equation*}

being $\nu_\perp$ the frequency of the parabolic trap, $V_0$ the depth of the shallow optical 
dipole potential, and $L$ the characteristic size. The potential barrier can be overcome, 
{\em without destroying it}, if there is a spatial variation of atomic interactions. 
Thus, we will consider the simple case in which the last term of eq. \ref{GPE} is modulated by 
a step function along $z$ of the form $\Delta(z<0)=0$ and $\Delta(z\geq 0)=U_2|\Psi|^2- U_3|\Psi|^4$ indicating
$U_2 = 4 \pi \hbar^2 a/m$ and $U_3 = U_2b$ the strength of two-body and three-body 
interactions, respectively. The previous energies are determined by the value of the s-wave scattering length $a$ 
and an effective volume parameter $b$, which indicates that three-body 
interactions act in a range $\approx b^{1/3}$. In the present work we will consider $a$ and $b$ as 
positive constants.

\section{One-dimensional eigenstates} 
We will consider setups in which the ground state of the optical dipole trap is much larger than the 
ground state of the transverse harmonic potential. In this situation we can describe the dynamics of 
the condensate in the quasi-one dimensional limit as given by a factorized wavefunction of the 
form \cite{Perezgarcia98,Jackson} $\Psi({\bf r},t)=\Phi_0(x,y)\cdot\chi(z,t)$. The density of atoms
per unit length is given by $|\chi|^2$. Normalizing the squared wavefunction
 by setting $|\psi|^2=\sqrt{8}\pi a|\chi|^2$, the following dimensionless equation 
is obtained:

\begin{equation*}
\label{NLSE}
i\frac{\partial \psi}{\partial \tau} = - \frac{1}{2}
\frac{\partial^2\psi}{\partial \eta^2} + f(\eta)\psi + \left(\alpha|\psi|^2- \beta |\psi|^4\right ) \psi,
\end{equation*}
where the normalized variables in time and space are 
$\tau=\nu_\perp t$ (time measured in units of the inverse of the radial trapping frequency)
and $\eta=z/r_\perp$ (length expressed in units of the transverse size of the cloud $r_\perp = \sqrt{\hbar/m\nu_\perp}$) 
where $f(\eta) = 2V_z/\hbar\nu_\perp$, $\alpha$ and $\beta=b/4\sqrt{2}ar_\perp^2$
are the effective trap, two-body and three-body atomic interaction
coefficients. eq. \ref{NLSE} is known as the cubic-quintic (CQ) nonlinear Schr\"odinger equation due to its 
dependency with the cubic and quintic powers of the wavefunction. This model has been extensively used in 
nonlinear optics \cite{nlo} as well as in superfluidity \cite{novoa09} and has known analytical solutions 
for the one-dimensional case.

\begin{figure}[htb]
{\centering \resizebox*{1\columnwidth}{!}{\includegraphics{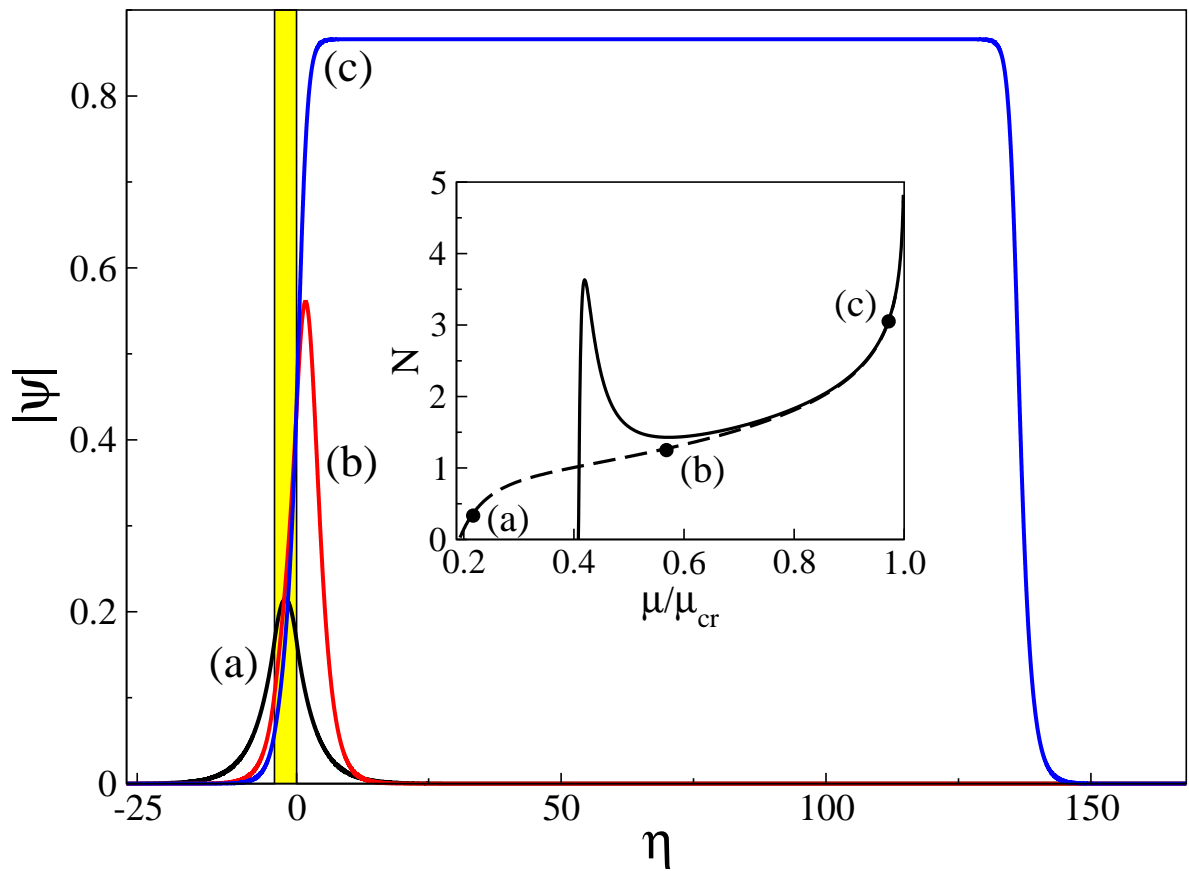}} \par}
\caption{ [Color Online] Amplitude profiles of several eigenstates of eq. \ref{NLSE} calculated for different 
numbers of atoms in the cloud. The right edge of the shallow trap (shaded zone) is located at $\eta=0$. 
For the zone with $\eta>0$, both attractive two-body and repulsive three-body interactions are characterized by 
$\alpha=\beta=1$. Inset: Dependence of the number of particles in the condensate $N$ on the chemical 
potential $\mu/\mu_{cr}$. The dashed (solid) line corresponds to the system displayed in the outer plot with a 
potential of width $L$ ($L'=4L$)). Labeled points refer to the eigenfunctions displayed.}   
\label{fig1}
\end{figure}
                                                                                                                        
In Fig. \ref{fig1} the geometry of the system to be considered in this
paper is shown. The shallow potential with depth $V_0$ and width $L$
is represented by the shaded region. Within the zone $\eta\geq 0$, non-zero interatomic 
interactions can be achieved by magnetically tuning the Feshbach resonances \cite{FB1,solitons2} 
or by optical manipulation \cite{FB2,zoller}.

We have searched numerically for eigenstates of eq. \ref{NLSE} with the form $\psi(\eta,\tau)=\phi(\eta)e^{i\mu \tau}$, 
where $\mu$ is the chemical potential. It is well-known that eq. \ref{NLSE} features a cut-off in the chemical 
potential spectrum, $\mu_{cr}=\frac{3\alpha^2}{16\beta}$, which constitutes the upper border of the existence domain 
of localized solitonic solutions \cite{novoa09}. 

Starting from the ground state of the trapping potential [(a) in Fig. \ref{fig1}] and by increasing gradually the 
number of atoms $N$, the eigenfunctions tend to enter the nonlinear zone [(b) in Fig.\ref{fig1}] displaying 
a high spatial localization. Once $N$ reaches a certain critical value the shape of the eigenstate approaches 
that of a continuous beam of fixed amplitude located outside the trap [(c) in Fig.\ref{fig1}]. In the latter case 
the value of the amplitude $A$ is completely determined by the strength of the atomic interactions and can be easily 
calculated for a plane wave solution of the CQ homogeneous model \cite{novoa09}.

As it can be appreciated in the inset of Fig.\ref{fig1}, in the case of a narrow trap of width $L$, the dashed curve 
$N$ vs. $\mu/\mu_{cr}$ monotonically increases (far enough from the critical value $\mu_{cr}$ where it becomes 
divergent) indicating that all eigenstates are stable. Otherwise, the solid curve corresponding to a system with a 
trap of width $L'=4L$ shows a region with a negative slope, yielding to the existence of an unstable domain. Notice 
that for $\mu\rightarrow\mu_{cr}$, both curves merge because the nonlinearity overcomes the trapping potential. 

\begin{figure}[htb]
{\centering \resizebox*{1\columnwidth}{!}{\includegraphics{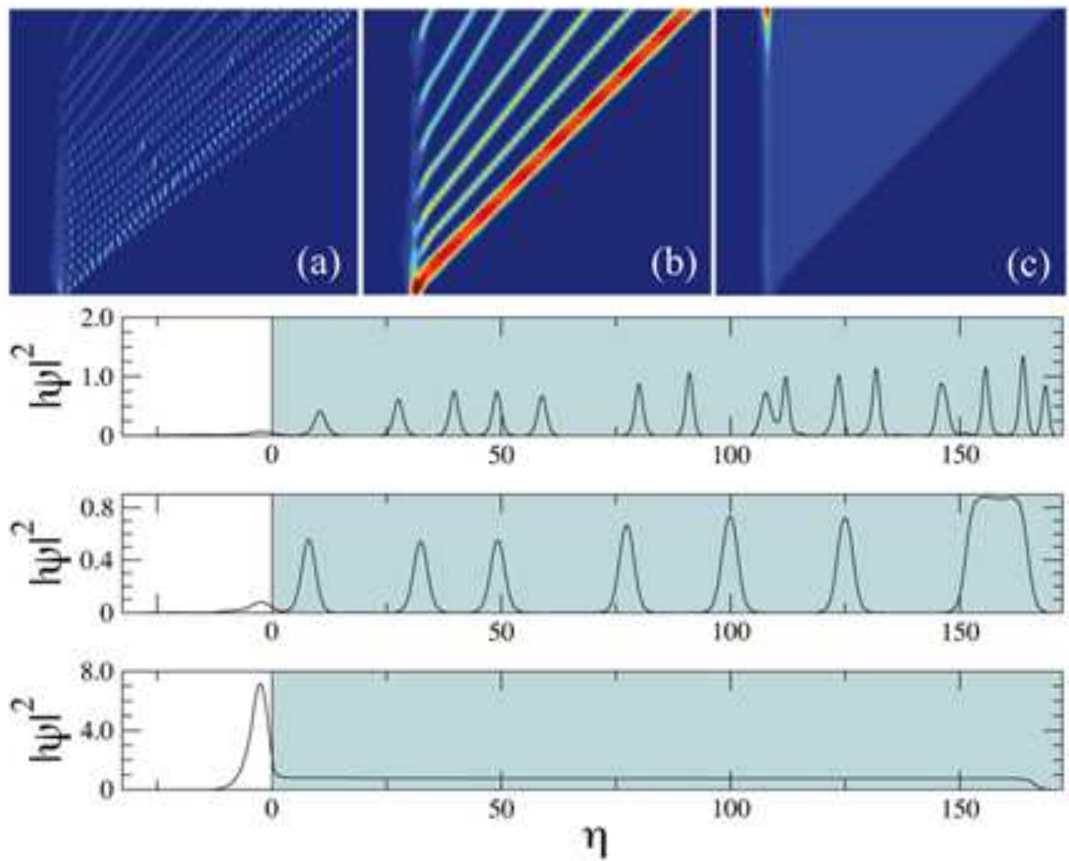}} \par}
\caption{[Color Online] Evolution of a Gaussian atomic cloud in the presence of localized gain 
$\Gamma=\Gamma_{cr}$ with two-body coefficient $\alpha=1$ and different values of the three-body atomic interactions: 
a) $\beta=0$. b) $\beta=0.8$. c) $\beta=1$. In all snapshots, the $x$-axis displays the spatial coordinate $\eta$ and
the $y$-axis corresponds to the evolution time $\tau\in[0,500]$. Below are the final density distribution plots 
of a) [top], b) [middle] and c) [bottom] figures.}
\label{fig2}
\end{figure}                                                                             

\section{Continuous wave emission} Taking into account the previous eigenstate structure of eq. \ref{NLSE}, 
we have performed a systematic analysis of the evolution dynamics of an initial Gaussian atomic cloud 
of unit amplitude and width $\omega=5$ placed at the center of a shallow square trap of depth $V_0=0.09$ and width $L=4$, 
by means of numerical simulations of eq. \ref{NLSE} in different parametric regimes. 
Subsequent feeding of the cloud with more atoms\cite{loading} has been simulated by adding a 
linear gain term $i\Gamma\psi$ to eq. \ref{NLSE} ($\Gamma\in\Re$) localized in the trapping region. 
In the absence of nonlinear interactions ($\alpha=\beta=0$) the gain effect is to increase the density of atoms 
in the trap. When only two-body interactions are considered ($\beta=0$) a burst of
solitons is emitted towards the region $\eta>0$ as predicted in previous work \cite{rodas-verde05} [see picture (a) in Fig. \ref{fig2}].
When $\beta>0$ and $\Gamma$ is set below a certain critical threshold $\Gamma_{cr}$ which depends on the parameters 
of the system (in the simulations of Fig. \ref{fig2}, we have fixed $\Gamma=0.0375$ which corresponds to $\Gamma_{cr}$ 
for $\alpha=\beta=1$), the first soliton emitted becomes wider featuring a flat-top profile \cite{novoa09} which is 
characteristic of systems with CQ nonlinearity [see picture (b) in Fig. \ref{fig2}]. However, if $\beta>0$ and 
$\Gamma\geq\Gamma_{cr}$ a continuous beam of constant amplitude is outcoupled from the trap 
[see picture (c) in Fig. \ref{fig2}]. Notice that the density plots in Fig. \ref{fig2} correspond to the final 
evolution stages ($\tau=500$) of simulations a)[top], b)[middle] and c)[bottom]. 

In terms of the physical description given above, it can be stated that whenever $\Gamma>0$ the system follows the 
eigenstate structure for growing values of $N$, but only above the critical value $\Gamma_{cr}$ the system is able 
to produce a coherent plane wave of constant amplitude $A$. Furthermore we have checked that the amplitude 
of the continuous matter wave released remains unaltered (unless small scale fluctuations) when three-body
losses \cite{Kagan} are taken into account.
We have numerically verified that whenever losses are moderate 
(typically less than $10^{-3}\beta$) their effect can be compensated by increasing the gain in the system.
For greater losses the analysis turns out to be more complicated and a more careful study, beyond the scope of 
this paper, must be performed.

\begin{figure}[htb]
{\centering \resizebox*{1\columnwidth}{!}{\includegraphics{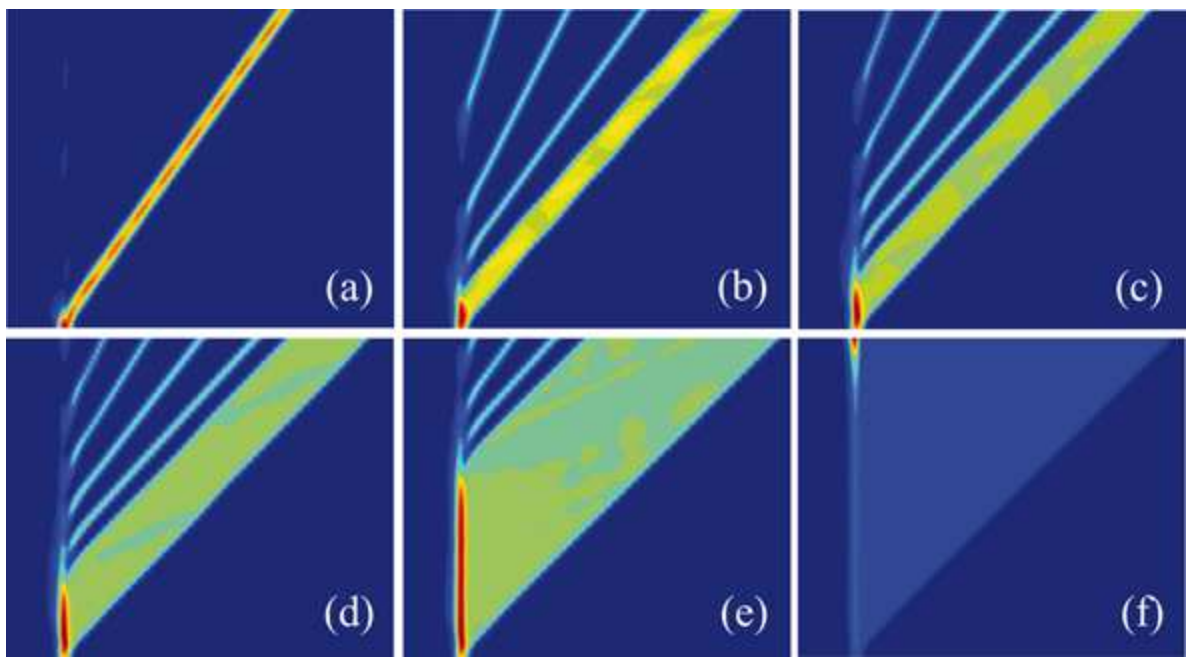}} \par}
\caption{\label{fig3}
[Color Online] Same as Fig. \ref{fig2} for different values of the gain coefficient $\Gamma/\Gamma_{cr}$. 
In all cases we have fixed $\alpha=\beta=1$. In a) $\Gamma/\Gamma_{cr}=0$; b) $\Gamma/\Gamma_{cr}=0.8$; c) $\Gamma/\Gamma_{cr}=0.93$; 
d) $\Gamma/\Gamma_{cr}=0.98$; e) $\Gamma/\Gamma_{cr}=0.99$; f) $\Gamma/\Gamma_{cr}=1$.}
\end{figure}



Another important feature of our system consists on the possibility of controlling the spatial width of the first 
soliton outcoupled. For a given pair of nonlinear coefficients [$\alpha,\beta$] this can be achieved by varying the gain 
coefficient $\Gamma$ as it can be seen if Fig. \ref{fig3}. In a) snapshot of Fig. \ref{fig3}, where $\Gamma$ is zero, almost 
all atoms are emitted in a coherent pulse of finite width. The increment in the spatial extension of the atomic soliton becomes 
non-negligible for small variations of the gain parameter in the regime $\Gamma/\Gamma_{cr}\approx 1$ [see pictures b) to f) 
in Fig. \ref{fig3}]. The case for which the rate of atoms loaded exactly compensates losses due to CW emission is 
illustrated in Fig. $3-f)$. It is noticeable that for $\Gamma>\Gamma_{cr}$, the CW emitted preserves its peak density and 
constant shape, indicating that the method is robust against perturbations. 

\begin{figure}[htb]
{\centering \resizebox*{1\columnwidth}{!}{\includegraphics{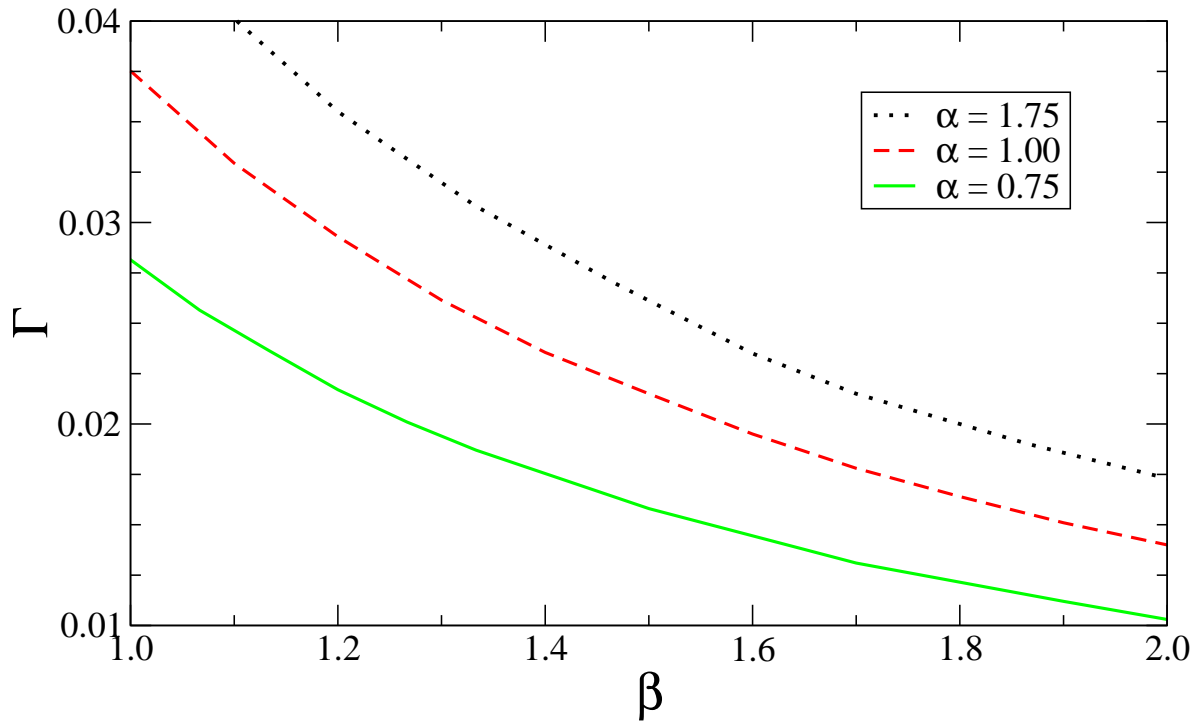}} \par}
\caption{\label{fig4}
[Color Online] Working range of the CW atomic laser for three different values of the two-body interaction 
parameter: $\alpha=1.75$[black pointed]; $\alpha=1$[red dashed] and $\alpha=0.75$[green continuous].
For each value of the three-body interaction coefficient $\beta$, it is plotted the corresponding $\Gamma_{cr}$ indicating that 
over this value CW emission is always achieved.}
\end{figure}

In Fig. \ref{fig4} we have plotted the results of many simulations in the whole parameter space, 
showing the working range of the CW atom laser for different $\alpha$ values. Plotted lines define 
the lower border of the operational region for each $\alpha$, i.e., line points correspond to $\Gamma=\Gamma_{cr}$ and 
CW emission is assured for gain values over such limit.
 
\section{Analytic condition for cw emission}
In the following we will derive an analytic expression for the continuous matter wave emission  based on waveguide 
modal theory \cite{Tamir}. 

We consider the linear trap (depicted in Fig. \ref{fig1})as a one-dimensional linear step-index waveguide 
with core index $n_c=V_0$, cladding index $n_{cl}=0$ (left trap border) and substrate index $n_s=V_{nl}$ 
(right border), being $V_{nl}$ the effective potential created by the nonlinear interactions. 
For the moment we will treat $V_{nl}$ as a constant value. In this context 
the guiding properties of the structure can be characterized by both the normalized frequency $f_0=L\sqrt{V_0-V_{nl}}$
and the cut frequency $f_c(\nu)=arctan{\sqrt{V_{nl}/(V_0-V_{nl})}}+\nu\pi$, where $\nu$ 
is a non-negative integer, which
establishes a cut-off for the existence of guided modes \cite{Tamir}. The specific case $f_c(0)$  corresponds to the 
threshold of the fundamental mode. We assume that if the lowest energy mode is cut there are not guided modes 
supported by the waveguide and just radiation modes can be excited. In other words, for $f_0<f_c(0)$
particles cannot be trapped and will eventually flow towards $\eta>0$.

Let us determine the values of $V_{nl}$ for which the non-guiding condition $F=f_0-f_c(0)< 0$ is fulfilled. 
Fixing $L$ and $V_0$ to the numbers employed in the simulations, we have found the limiting value
$V_{nl}^c=0.05$ numerically solving $F=0$. Thus for $V_{nl}>V_{nl}^c$ the guide cannot retain the matter wave field, while in the opposite case  the particles will remain trapped if the proper mode is correctly excited. 
Hereafter we will consider 
$V_{nl}$ to be field-dependent owing to the existence of nonlinear interactions within the region $\eta\geq0$.
On the right border of the dipole trap, $V_{nl}^c=\alpha|\psi|^2-\beta|\psi|^4$ indicating that the potential barrier 
depends on the particle density. Notice that as  $|\psi|^2$ must be real the latter expression predicts a 
cut-off in the outcoupling mechanism since for $V_{nl}^c\geq\frac{\alpha^2}{4\beta}$ the emission is forbidden. 
The agreement between analytical and numerical estimations of this cut-off is above $90\%$.

We have also determined the dependence of $V_{nl}^c$ on $V_0$  (fixing $L$) to be linear with 
slope $\xi=0.917$, ordinate $\chi=-0.039$ and  correlation $R^2=0.9998$.
Finally, assuming that the atomic cloud features a Gaussian transverse profile as in the whole simulations
we arrive at the analytic expression for the emission condition:

\begin{equation*}
A^2\geq\frac{\alpha+\sqrt{\alpha^{2}-4\beta(\xi V_0+\chi)}}{2\beta e^{-\frac{L^2}{2\omega^2}}};
\end{equation*}

where A is the amplitude of the Gaussian atomic cloud loaded in the trap. This condition establishes a 
lower limit for matter wave emission. Continuous operation of the atom laser will be assured 
by keeping the peak particle density $A^2$ above this threshold. In our numerical simulations we 
have added a linear gain factor in order to satisfy this condition. Experimentally this could be realized
by continuously  loading the optical trap with a flux of particles \cite{loading}.

\section{Discussion and conclusions} We have shown in this work that
CW operation of a matter-wave laser could be realized within an ample
parameter space for negative scattering length and repulsive three-body elastic 
interactions. Although the cubic-quintic model can be considered an exotic system, 
it has been proposed that {\em repulsive} three-body forces can 
be activated by simultaneously switching on an optical lattice in $\eta\geq 0$ 
containing an ultracold gas of particles with high magnetic or electric 
dipole moment \cite{zoller}.

In summary, we have proposed a novel mechanism for continuous operation of a coherent matter wave laser. 
Our system is able to perform a regular and controllable emission of coherent atomic beams 
of constant amplitude with the only limitation of the number of particles that can be loaded in the trap. 
It is important to note that even though the present paper is devoted to square traps, our results also apply 
to other trap configurations without loss of generality. We have also shown how the width of the first 
atomic soliton released can be easily controlled by means of variations of the input gain. 
Moreover we have derived an analytic condition for the CW emission which relates the peak particle density
of the atomic cloud with the tunable parameters of the system. 
As the techniques for coherently feeding BECs progress our idea could provide a novel approach to new 
types of matter-wave lasers.

\section{Acknowledgments}
This work was supported by Ministerio de Educaci\'on y Ciencia, Spain
(project FIS2008-01001) and University of Vigo (project 08VIA09).  D. Novoa acknowledges support from 
Conseller\'{i}a de Innovaci\'on e Industria-Xunta de Galicia through the {\it Mar\'{i}a Barbeito} program.


\end{document}